\documentstyle[seceq,preprint,epsbox]{jpsj}

\def\runtitle{
Resonant X-Ray Magnetic Scattering from CoO
}
\def\runauthor{Jun-ichi {\sc Igarashi} and Manabu {\sc Takahashi}}

\title{Resonant X-Ray Magnetic Scattering from CoO}

\author{Jun-ichi {\sc Igarashi} and Manabu {\sc Takahashi}}

\inst{Faculty of Engineering, Gunma University, Kiryu, Gunma 376-8515}

\recdate{
\ \ \ \ \ \ \ \ \ \ \ \ \ \ \ 
}

\def\vol(#1,#2,#3){{\bf #1} (#2) #3}
\abst
{
We analyze the recent experiment [W. Neubeck {\em et al.},
Phys. Rev. B \vol(60,1999,R9912)] for the resonant x-ray magnetic 
scattering (RXMS) around the K edge of Co in the antiferromagnet CoO.
We propose a mechanism of the RXMS to make the $4p$ states couple 
to the magnetic order:
the intraatomic exchange interaction between the $4p$ and
the $3d$ states and the $p$-$d$ mixing to the $3d$ states
of neighboring Co atoms. 
These couplings induce the orbital moment in the $4p$ states
and make the scattering tensor antisymmetric.
Using a cluster model, we demonstrate that this modification 
gives rise to a large RXMS intensity in the dipole process,
in good agreement with the experiment.
We also find that the pre-edge peak is generated by the transition
to the $3d$ states in the quadrupole process,
with negligible contribution of the dipole process.
We also discuss the azimuthal angle dependence of the intensity.
} 

\kword
{
resonant x-ray scattering, magnetic scattering, $K$-edge,
CoO, quadrupole transition}
 
\begin{document}
\sloppy
\maketitle
\section{Introduction}

Resonant x-ray scattering has recently attracted much interest
as a useful tool to investigate the magnetic order.
The resonant enhancement for magnetic Bragg 
reflections has been observed in transition metals\cite{Namikawa} 
and their compounds.\cite{Hill,Neubeck,Stunault}
For the resonant enhancement to be observable on the magnetic Bragg spots,
orbitals have to couple with the magnetic order. It is usually achieved
through the spin-orbit interaction.
In this paper, we study the mechanism of the resonant x-ray magnetic 
scattering (RXMS) by focusing our attention on a typical material, CoO,
where orbitals are strongly coupled to the magnetic order, 
and analyze the recent RXMS experiment by Neubeck {\em et al}.\cite{Neubeck}

\begin{figure}
\begin{center}
\leavevmode
\epsfile{file=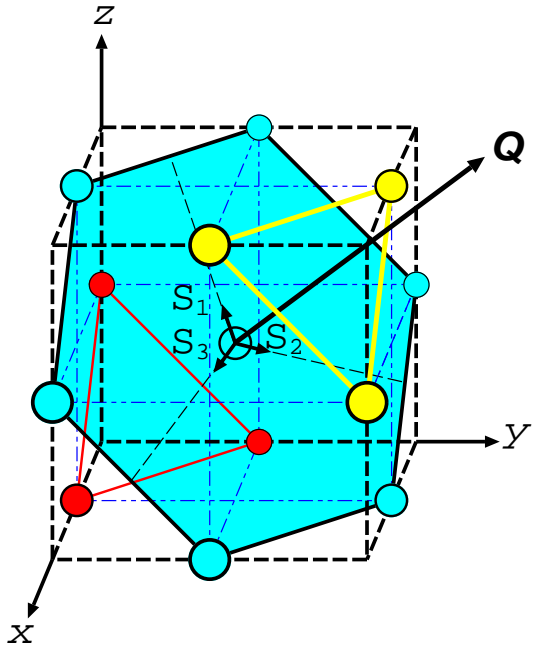,width=7cm}
\caption{
Sketch of a CoO crystal (only Co atoms are shown).
Wave vector ${\bf Q}=(\frac{1}{2},\frac{1}{2},\frac{1}{2})$
characterizes an antiferromagnetic modulation direction.
Three arrows in a plane perpendicular to ${\bf Q}$ indicate
possible spin directions.
Spins of solid, gray, and open circles align alternately.
}\label{fig.cryst}
\end{center}
\end{figure}

In CoO, Co atoms form an fcc lattice with a small tetragonal distortion 
of $c/a<1$ below the N\'eel temperature $T_N=292$ K. 
It is a type II antiferromagnet (AF) characterized by a wavevector
${\bf Q}$, which is directed to one of four body-diagonals 
in the fcc lattice, thus forming $K$ domains.\cite{Shull,Ronzaud}
Although a non-collinear magnetic order has been proposed so far,
\cite{vanLaar,Shishido} it is quite reasonable to assume a collinear magnetic
order in the analysis of the RXMS intensity, since
the RXMS signal is separately observed for different domains
in the experiment.\cite{Neubeck}
The spin vector lies in the plane perpendicular 
to the AF modulation direction; three directions are possible,
forming $S$ domains.\cite{Ronzaud} 
For an example, the spins are directed to $(-1,-1,2)$, $(-1,2,-1)$, 
or $(2,-1,-1)$, which we call domains $S_1$, $S_2$, and $S_3$, 
in the $K$ domain with ${\bf Q}=(\frac{1}{2},\frac{1}{2},\frac{1}{2})$.
Figure \ref{fig.cryst} depicts schematically this domain.
In total, we have 12 domains. 
The orbital moment is known to be as large as $\sim 1\mu_B$.
\cite{Shull} This material was expensively studied for decades ago 
by Kanamori,\cite{Kanamori} who pointed out that the large orbital moment 
causes a large magnetostriction. 

\begin{figure}
\begin{center}
\leavevmode
\epsfile{file=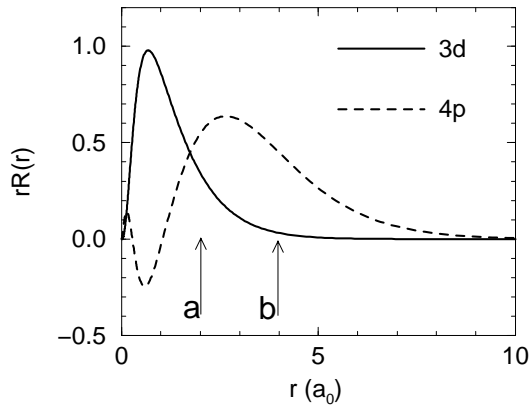,width=7cm}
\caption{
Radial wave functions (multiplied by the radius) in the $3d^7$ 
configuration of a free Co$^{2+}$ atom. The radius $r$ is measured 
in units of the Bohr radius $a_0$.
The arrow $a$ indicates the ion radius, which is defined by
the half of the distance between the Co and O atoms in the CoO crystal.
The arrow $b$ indicates the distance to a neighboring O atom.
}\label{fig.atom}
\end{center}
\end{figure}

For photons with energy near the $K$ edge,
the $4p$ states become occupied in the intermediate states of 
the dipole process. The $4p$ states, different from the $3d$ states,
are highly extended in space. To have a feeling of the character of
the wavefunctions, we show in Fig.~\ref{fig.atom}
the radial wave functions calculated within the Hartree-Fock approximation
\cite{Cowan} for a free Co$^{2+}$ atom.
The concept of localized orbitals loses its clear meaning outside the ion 
radius in solids. Since the $4p$ states have large weights outside
the ion radius, treating them as local orbitals in a cluster model 
may limit the usefulness of the model to a qualitative level.
Nevertheless, we use a cluster model to make clear the mechanism of 
the RXMS in CoO. We hope this will serve as a first step to
more quantitative study based on band structure calculations.

A closely related phenomenon where the $4p$ states are involved
is the resonant x-ray scattering for the orbital order in LaMnO$_3$.
\cite{Murakami,Ishihara} This case has been thoroughly studied 
with treating the $4p$ states of Mn as a band.
\cite{Elfimov,Benfatto,Takahashi1}
It has been shown that the $4p$ states of Mn is mainly modified 
by the Jahn-Teller distortion via the oxygen potential on the neighboring 
sites, and that this modification gives rise to non-vanishing diagonal 
elements in the scattering tensor, leading to a large main peak.
On the other hand, the mechanism of the RXMS for CoO is much complicated.
The $4p$ states are required to couple to the magnetic order.
This is achieved through the coupling to the $3d$ states which constitute 
the magnetic order. Two processes are likely to be effective to make
the $4p$ states couple to the $3d$ states: 
the intraatomic exchange interaction and the $p$-$d$ mixing 
to the $3d$ states of neighboring Co atoms.
We demonstrate that these couplings give rise to a modification 
of the $4p$ states and make the scattering tensor antisymmetric.
The RXMS intensity thus generated seems much smaller than the intensity
generated by the Jahn-Teller effect in LaMnO$_3$,
but is sufficiently larger than the pre-edge intensity coming from
the quadrupole process. Note that
the $3d$ states having orbital moment is crucial to generate the RXMS
intensity. The amplitude of exciting an electron with up spin 
(up spin process) is generally different from that of the exciting 
an electron with down spin (down spin process). 
However, the sum of the amplitudes for
the up and down spin processes become the same at each site
if the system has no orbital moment;
the RXMS intensity is not generated by such spin scattering processes alone, 
since the total amplitude is given by the difference between those 
at the $A$ and $B$ sublattices.

In addition to the dipole process, we also consider the quadrupole process,
which gives rise to a pre-edge peak.
Although the dipole process can also contribute to the pre-edge peak through
the $p$-symmetric states mixing to the $3d$ states of neighboring Co atoms, 
this contribution is found to be much smaller than
that of the quadrupole process.
This result is consistent with the experiment.\cite{Neubeck}
Note that, for the pre-edge peak in LaMnO$_3$, the contribution from 
the dipole process is predicted to be much larger than that of the
quadrupole process.\cite{Takahashi2} 

This paper is organized as follows. In \S~2, we formulate
the RXMS intensity. In \S~3, we present
a cluster model and discuss the relevant $4p$ and $3d$ states.
In \S~4, we show the calculated results for the RXMS intensity 
in comparison with the experiment.
Section 5 is devoted to concluding remarks.
In the Appendix, the geometrical factor is explicitly given 
for the dipole and quadrupole processes.

\section{X-Ray Magnetic Scattering}

\begin{figure}
\begin{center}
leavevmode
\epsfile{file=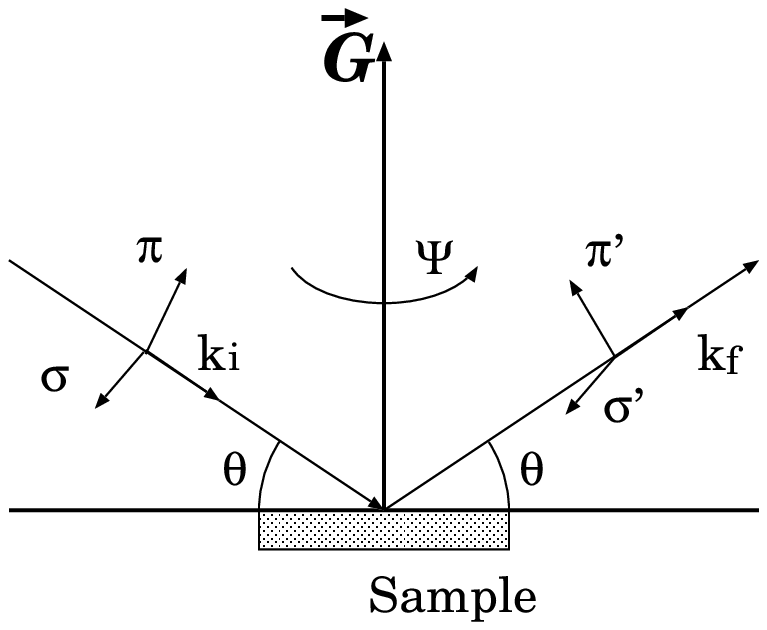,width=5cm}
\caption{
Geometry of x ray scattering. Photon with polarization
$\sigma$ or $\pi$ is scattered into the state of polarization
$\sigma'$ or $\pi'$ at Bragg angle $\theta$.
The sample crystal is rotated by azimuthal angle $\psi$
around the scattering vector ${\bf G}={\bf k}_f-{\bf k}_i$.
}\label{fig.geom}
\end{center}
\end{figure}

The magnetic scattering amplitude has been reviewed by several authors.
\cite{deBergevin1,Blume1,Blume2}
We summarize here the expression of the amplitude in a form 
suitable to calculating the azimuthal angle dependence.
The scattering geometry is shown in Fig.~\ref{fig.geom}.
Photon with frequency $\omega$, momentum ${\bf k}_i$ and polarization $\mu$ 
($=\sigma$ or $\pi$) is scattered into the state
with momentum ${\bf k}_f$ and polarization $\mu'$ ($=\sigma'$ or $\pi'$). 
For the system with two magnetic sublattices $A$ and $B$ in the AF order,
we write down the cross section for elastic scattering
at magnetic Bragg peaks as
\begin{equation}
 \left. \frac{d\sigma}{d\Omega'}\right|_{\mu\to\mu'} \propto
  \left| T^A_{\mu\to\mu'}({\bf G},\omega) 
       - T^B_{\mu\to\mu'}({\bf G},\omega) \right|^2 ,
\end{equation}
with
\begin{eqnarray}
  T^{\eta}_{\mu\to\mu'}({\bf G},\omega) =
    &-&\frac{i\hbar\omega}{mc^2}\left(\frac{1}{2}{\bf L}^{\eta}({\bf G})
      \cdot {\bf A}''+{\bf S}^{\eta}({\bf G})\cdot{\bf B}\right)\nonumber\\
    &+&J^{\eta}_{\mu\to\mu'}({\bf G},\omega)
     +L^{\eta}_{\mu\to\mu'}({\bf G},\omega).
\label{eq.amp}
\end{eqnarray}
where ${\bf G}={\bf k}_f-{\bf k}_i$ is the scattering vector.
The first term in eq.~(\ref{eq.amp}) represents a {\em non-resonant}
term at sublattice $\eta$ ($=A$ or $B$),
with $m$ and $c$ being the electron mass and the velocity of photon, 
respectively.\cite{Blume2}
${\bf L}^\eta ({\bf G})$ and ${\bf S}^\eta({\bf G})$ stand for the Fourier
transform of the orbital and spin angular momentum density,\cite{com1}
and ${\bf A}''$ and ${\bf B}$ are given by
\begin{eqnarray}
 {\bf A}''&=& {\bf A}'-({\bf A}'\cdot{\bf\hat G}){\bf\hat G},
  \quad {\bf A}'=-4\sin^2\theta(\hat{\mbox{\boldmath $\epsilon$}}'\times
  \hat{\mbox{\boldmath $\epsilon$}}),\\
 {\bf B} &=& \hat{\mbox{\boldmath $\epsilon$}}'\times 
 \hat{\mbox{\boldmath $\epsilon$}}
             + ({\bf\hat k}_f\times \hat{\mbox{\boldmath $\epsilon$}}')
               ({\bf\hat k}_f\cdot \hat{\mbox{\boldmath $\epsilon$}})
             - ({\bf\hat k}_i\times \hat{\mbox{\boldmath $\epsilon$}})
               ({\bf\hat k}_i\cdot \hat{\mbox{\boldmath $\epsilon$}}')
             - ({\bf\hat k}_f\times \hat{\mbox{\boldmath $\epsilon$}}')
        \times ({\bf\hat k}_i\times \hat{\mbox{\boldmath $\epsilon$}}),
\end{eqnarray}
where $\hat{\mbox{\boldmath $\epsilon$}}$ and 
$\hat{\mbox{\boldmath $\epsilon$}}'$ are the initial
and scattered polarizations, and
${\bf\hat k}_i$, ${\bf\hat k}_f$, and ${\bf\hat G}$ are normalized 
vectors of ${\bf k}_i$, ${\bf k}_f$, and ${\bf G}$.

The second and third terms in eq.~(\ref{eq.amp}) represent 
the resonant contribution at the $K$ edge, which originates from
a second-order process that a photon is virtually absorbed 
by exciting the $1s$ electron to unoccupied states and then emitted
by recombining the excited electron with the $1s$ core hole.
The amplitude by the dipole process is expressed as 
\begin{equation}
  J_{\mu\to\mu'}^\eta({\bf G},\omega)
  = \sum_{\alpha\alpha'}(P'^{\mu'})^\dagger_{\alpha}
    M_{\alpha\alpha'}^\eta(\omega)P^{\mu}_{\alpha'},
\label{eq.damp}
\end{equation}
where the geometrical factors $P^\mu$, $P'^{\mu'}$ are explicitly given in
the Appendix, and the scattering tensor $M^\eta(\omega)$ is given by
\begin{equation}
 M_{\alpha\alpha'}^\eta(\omega) = \sum_j 
  \frac{m\omega_{jg}^2\langle g|x_\alpha|j\rangle\langle j|x_{\alpha'}|g\rangle}
       {\hbar\omega-(E_j-E_g)+i\Gamma/2}.
\label{eq.dipole}
\end{equation}
Here $x_\alpha$'s stand for the dipole operators, $x_1=x$, $x_2=y$,
and $x_3=z$ in the coordinate frame fixed to the crystal axes. 
The ground state $|g\rangle$ has an energy $E_g$,
and the intermediate state $|j\rangle$, consisting of an excited electron 
on the $4p$ states and a hole on the $1s$ state, has an energy $E_j$;
$\omega_{jg}\equiv (E_j-E_g)/\hbar$.
Quantity $\Gamma$ is introduced to describe the life-time broadening
due to the core hole. 
The amplitude by the quadrupole process is expressed as 
\begin{equation}
 L_{\mu\to\mu'}^\eta({\bf G},\omega)
  = \sum_{\gamma\gamma'}(Q'^{\mu'})^\dagger_{\gamma}
    N_{\gamma\gamma'}^\eta(\omega)Q^{\mu}_{\gamma'},
\label{eq.qamp}
\end{equation}
where the geometrical factors, $Q^\mu$, $Q'^{\mu'}$, are explicitly
given in the Appendix, and the scattering tensor $N^\eta(\omega)$ is given
by
\begin{equation}
 N_{\gamma\gamma'}^\eta(\omega) = \frac{k^2}{12}\sum_{\ell}
  \frac{m\omega_{\ell g}^2\langle g|z_\gamma|\ell\rangle\langle\ell
       |z_{\gamma'}|g\rangle}
       {\hbar\omega-(E_\ell-E_g)+i\Gamma/2}.
\label{eq.quadrupole}
\end{equation}
Here $k$ is the wavenumber of the incident (and scattered) photon,
and $z_\mu$'s stand for the quadrupole operators,
$z_1=(\sqrt{3}/2)(x^2-y^2)$, $z_2=(1/2)(3z^2-r^2)$, $z_3=\sqrt{3}yz$,
$z_4=\sqrt{3}zx$, and $z_5=\sqrt{3}xy$ in the coordinate frame
fixed to the crystal axes.
The intermediate state $|\ell\rangle$, consisting of an excited electron 
on the $3d$ states and a hole on the $1s$ states,
has an energy $E_\ell$; $\omega_{\ell g}=(E_\ell - E_g)/\hbar$.

\section{$3d$ and $4p$ States at Co Sites}

Each Co atom is in the $3d^7$ configuration under the cubic crystal field.
We neglect small effects of the distortion such that $c/a < 1$.
The ground state $^4F$ of a free 
cobaltous ion is split into the representations $\Gamma_4$ and $\Gamma_5$.
Three degenerate states in the $\Gamma_4$ representation have lower
energy than those in the $\Gamma_5$ representation.
The spin of each ion is subjected to a strong exchange
field and is oriented in its direction, forming an AF
long range order. As mentioned before, we assume that the AF modulation 
direction ${\bf Q}$ is one of the four body-diagonals in the fcc lattice.

As regards the orbitals, we introduce the pseudo-angular-momentum 
operator $\mbox{\boldmath$\ell$}$ of magnitude 1 in the $\Gamma_4$ 
representation, such that the orbital-angular-momentum operator 
${\bf L}$ is given by $-(3/2)\mbox{\boldmath$\ell$}$.
Thereby, in the first approximation, the wave function of 
the state specified by $\ell_z=0$ correspond to
the configuration that $\varphi_0\equiv d_{xy}$ is singly occupied
(in other wards, $d_{zx}$ and $d_{yz}$ are doubly occupied);
the state $\ell_z=\pm 1$ is such that $\varphi_{\pm 1}\equiv
=\mp\frac{1}{\sqrt{2}}(d_{zx}\pm id_{yz})$ is singly occupied.
Connected with the spin through the spin-orbit coupling, 
\mbox{\boldmath$\ell$} becomes antiparallel to the spin direction. 
Thus the ground state gives the spin moment $\sim 3\mu_B$
and the orbital moment $\sim 1.5\mu_B$.
As discussed by Kanamori,\cite{Kanamori} quantum mechanical effects 
modify this classical ground state, making the magnetic moment close
to the experimental value. Since we are aiming at 
analyzing qualitatively the RXMS intensity,
we will not consider the quantum correction.

The unoccupied $3d$ states become occupied by an electron excited 
from the $1s$ state in the intermediate states of the quadrupole process.
The scattering amplitudes of filling the unoccupied $3d$ states
are usually cancelled out between those at the $A$ and $B$ sublattices.
However, it is not the case for the unoccupied state in the $\Gamma_4$ 
representation, since this orbital is strongly coupled
to the magnetic order through the spin-orbit coupling.
We call this unoccupied state in the intermediate state as the $3d^*$ 
state with energy $\epsilon_d$. This $3d^*$ state, which is represented 
as $\varphi'_{\mp 1}$ in the coordinate frame that the $z$ axis is 
along the spin direction (the upper (lower) sign corresponds to the $A$ ($B$) 
sublattice), is expressed in the coordinate frame fixed to the crystal 
axes as
\begin{equation}
  \varphi'_{\mp 1}= D^{(1)}_{1\mp 1}(\alpha,\beta,\gamma)\varphi_1
                   +D^{(1)}_{0\mp 1}(\alpha,\beta,\gamma)\varphi_0
                   +D^{(1)}_{-1\mp 1}(\alpha,\beta,\gamma)\varphi_{-1},
\label{eq.single}
\end{equation}
where $D^{(1)}_{mm'}(\alpha,\beta,\gamma)$ is the rotation matrix
of the first rank,\cite{Rose} with the Euler angles 
$\alpha=-3\pi/4$, $\beta=\arcsin{\frac{1}{\sqrt{3}}}$, $\gamma=0$
for domain $S_1$, $\alpha=\pi/2+\arcsin{1/\sqrt{5}}$,
$\beta=\pi-\arcsin\sqrt{5/6}$, $\gamma=0$ for domain $S_2$,
and $\alpha=-\arcsin{1/\sqrt{5}}$,
$\beta=\pi-\arcsin\sqrt{5/6}$, $\gamma=0$ for domain $S_3$.
Substituting these states for $|\ell\rangle$ in eq.~(\ref{eq.quadrupole}),
we obtain the scattering tensor, which has an antisymmetric form,
\begin{equation}
 N^A(\omega)-N^B(\omega)
   = \left( \begin{array}{rrrrr}
            0 & 0 & 0 & 0 & 0 \\
            0 & 0 & 0 & 0 & 0 \\
            0 & 0 & 0 & d & f \\
            0 & 0 &-d & 0 & e \\
            0 & 0 &-f &-e & 0
            \end{array} \right) ,
\end{equation}
with $d$, $e$, and $f$ being complex numbers.

The unoccupied $4p$ states at Co sites become occupied 
by an electron excited from the $1s$ state in the intermediate states
of the dipole process.
The scattering amplitudes of filling the $4p$ states are 
cancelled out between those at the $A$ and $B$ sublattices, 
unless the $4p$ states are coupled to the $3d$ states which constitute 
the magnetic order. 
One possible coupling is the intraatomic exchange interaction
between the $4p$ and the $3d$ states.
Since the occupied $3d$ states have finite orbital moment,
this coupling makes the $4p$ levels split; 
at the $A$ sublattice,
the states
$p'_1=-\frac{1}{\sqrt{2}}(p'_x+ip'_y)$, $p'_0=p'_z$ and
$p'_{-1}=\frac{1}{\sqrt{2}}(p'_x-ip'_y)$ with down spin have energy 
$\epsilon_1$, $0$, $\epsilon_2$; 
at the $B$ sublattice, $p'_1$, $p'_0$, $p'_{-1}$ with up spin 
have energy $\epsilon_2$, $0$, $\epsilon_1$.
Here $p'_{\pm 1}$ and $p'_0$ refer to
the coordinate frame that the $z$ axis is along the spin direction.
The $\epsilon_1$ and $\epsilon_2$ are roughly evaluated as
$\epsilon_1=-0.056$ eV, $\epsilon_2=-0.020$ eV.\cite{com2}
Another coupling is the mixing to the $3d$ states of neighboring Co atoms.
What is effective to make the amplitude at the $A$ sublattice
different from at the $B$ sublattice is the mixing to the $3d^*$ state.

To estimate the effects of these couplings, we use a cluster model that
the $4p$ states at the central site are mixing to the $3d^*$ state
of 12 neighboring Co atoms (Fig.~\ref{fig.cryst}).
Thus the Hamiltonian matrix has $39\times 39$ dimensions.
The effect of the mixing to the $2p$ states of neighboring oxygens 
is implicitly included into the model by adjusting the energy of 
the $4p$ states. According to the experiment,\cite{Neubeck} 
the main peak is at $\hbar\omega=7.724$ keV and 
the pre-edge peak at $\hbar\omega=7.707$ keV.
Thereby we put $\epsilon_p-\epsilon_d=17$ eV for the difference 
between the energy of the $4p$ states, $\epsilon_p$, and that of 
the $3d^*$ states, $\epsilon_d$, both at the central site.
Note that
both $\epsilon_p$ and $\epsilon_d$ are including the interaction energy
with the $1s$ hole. 
The energy $\tilde\epsilon_d$ of the $3d^*$ state at the neighboring Co 
atoms is several eV's higher than that at the central site 
where the core hole exists,
since the interaction with the $1s$ hole is not working there.
We assume the difference to be $\tilde\epsilon_d-\epsilon_d=4$ eV.
Then we evaluate the off-diagonal part of the Hamiltonian matrix 
by using the Slater-Koster parameters, $(pd\sigma)$ and 
$(pd\pi)$,\cite{Slater} with the help of eq.~(\ref{eq.single}).
We assume $(pd\sigma)=-0.5$ eV together with an empirical relation 
$(pd\pi)=-(1/2)(pd\sigma)$.\cite{Mattheiss}
This size of the $(pd\sigma)$ value seems reasonable, since
the $4p$ states are highly extended in space.

The Hamiltonian matrix can be numerically diagonalized.
We find that three degenerate $4p$ states are 
split into the states with finite orbital moments.
Substituting these eigenstates for the intermediate states 
in eq.~(\ref{eq.dipole}), we calculate the scattering amplitude. 
As a reflection of the induced orbital moment for the $4p$ states,
the scattering tensor has an antisymmetric form, 
\begin{equation}
 M^A(\omega)-M^B(\omega)
   = \left( \begin{array}{rrr}
            0 & a & c \\
           -a & 0 & b \\
           -c &-b & 0
            \end{array} \right) ,
\end{equation}
with $a$, $b$, and $c$ being complex numbers.
 
\section{Calculated Results in Comparison with Experiment}

In the preceding section, we have studied the $3d$ and $4p$ states 
in the intermediate states of the RXMS. 
For calculating the scattering intensity,
we have to estimate further several quantities.
First, we evaluate the matrix elements for the dipole and
quadrupole processes within the HF approximation.\cite{Cowan}
They are given by
\begin{eqnarray}
 \langle 4p|r|1s\rangle &\equiv&
   \int_0^{\infty} R_{4p}(r)rR_{1s}(r)r^2{\rm d}r = 1.52\times 10^{-11} 
   \, {\rm cm}, \\
 \langle 3d|r^2|1s\rangle &\equiv&
   \int_0^{\infty} R_{3d}(r)r^2R_{1s}(r)r^2{\rm d}r = 2.10\times 10^{-20} 
   \, {\rm cm}^2,
\end{eqnarray}
where $R_{1s}(r)$, $R_{4p}(r)$, and $R_{3d}(r)$ are radial wave functions 
for the $1s$, $4p$, and $3d$ states in the $3d^7$ configuration
of a free cobaltous ion (see Fig.~\ref{fig.atom}). 
Since the $1s$ state is well localized inside the ion radius,
these values are to be good estimates in solids, with multiplying 
a factor $0.8$.
With these values and $k\sim 3.9\times 10^8$ cm$^{-1}$ around the $K$ edge, 
we have $k^2|\langle 3d|r^2|1s\rangle|^2/(20|\langle 4p|r|1s\rangle|^2)
\sim 1.5\times 10^{-2}$, 
which is a measure of the ratio of the amplitude 
of the quadrupole process to that of the dipole process.
Second, we assume that $|{\bf S}({\bf G})|\sim 1.0$,
$|{\bf L}({\bf G})|\sim 0.8$, 
in the first term of eq.~(2.3), 
for the scattering vectors ${\bf G}=(-\frac{1}{2},-\frac{1}{2},\frac{7}{2})$ 
and $(\frac{1}{2},\frac{1}{2},\frac{5}{2})$; 
these values are smaller than those for ${\bf G}=0$, that is, 
the local orbital- and spin-angular momenta. 
This is consistent with the general tendency that 
$|{\bf S}({\bf G})|$ and $|{\bf L}({\bf G})|$ are decreasing with
increasing values of ${\bf G}$.\cite{Fernandez} 
The precise values are not necessary for our qualitative analysis.
Third, we use the experimental value $1.7$ eV from the fluorescence 
measurement for the life-time broadening width $\Gamma$.\cite{deBergevin2}

Experimentally, Neubeck {\em et al.} have separated 
the contributions to the scattering intensity from each magnetic domain;
the data are shown for ${\bf G}=(-\frac{1}{2},\frac{1}{2},\frac{7}{2})$ 
and $(\frac{1}{2},-\frac{1}{2},\frac{5}{2})$ 
in the $K$ domain with ${\bf Q}=(\frac{1}{2},-\frac{1}{2},\frac{1}{2})$.
\cite{Neubeck}
In this paper, we consider the $K$ domain with 
${\bf Q}=(\frac{1}{2},\frac{1}{2},\frac{1}{2})$, not identical to 
the experimental one. In our $K$ domain, the scattering vectors
${\bf G}=(-\frac{1}{2},-\frac{1}{2},\frac{7}{2})$
and $(\frac{1}{2},\frac{1}{2},\frac{5}{2})$ correspond to the above
experimental scattering vectors.
Note that the $(1,-1,0)$ crystal axis lies on the scattering 
plane at $\psi=0$ in our definition of the geometrical
factor (see the Appendix).

Figure \ref{fig.spec1} (a) shows the calculated intensity 
as a function of photon energy 
for ${\bf G}=(-\frac{1}{2},-\frac{1}{2},\frac{7}{2})$.
The azimuthal angle is set to be $\psi=-46^\circ$, such that
the (1,0,0) crystal axis lies on the scattering plane.
(This corresponds to the experimental geometry 
that the $(0,1,0)$ crystal axis lies on the scattering plane.)
For the $\sigma\to\sigma'$ channel (lower panel), only one peak 
is found at the pre-$K$-edge position, in agreement with the experiment. 
This peak arises from the quadrupole process, 
since the dipole process is forbidden. 
The amplitude from the quadrupole process interferes with the non-resonant 
term, leading to an anti-resonant dip in the spectra.
For the $\sigma\to\pi'$ channel (upper panel), 
a large peak is found at the main $K$-edge position
in addition to the pre-edge peak, 
in good agreement with the experimental curve after correction for 
absorption.\cite{Neubeck,com3}
The intensity of the main peak depends on the $S$ domains;
for example, it is very small for domain $S_3$.
\begin{figure}
\begin{center}
\leavevmode
\epsfile{file=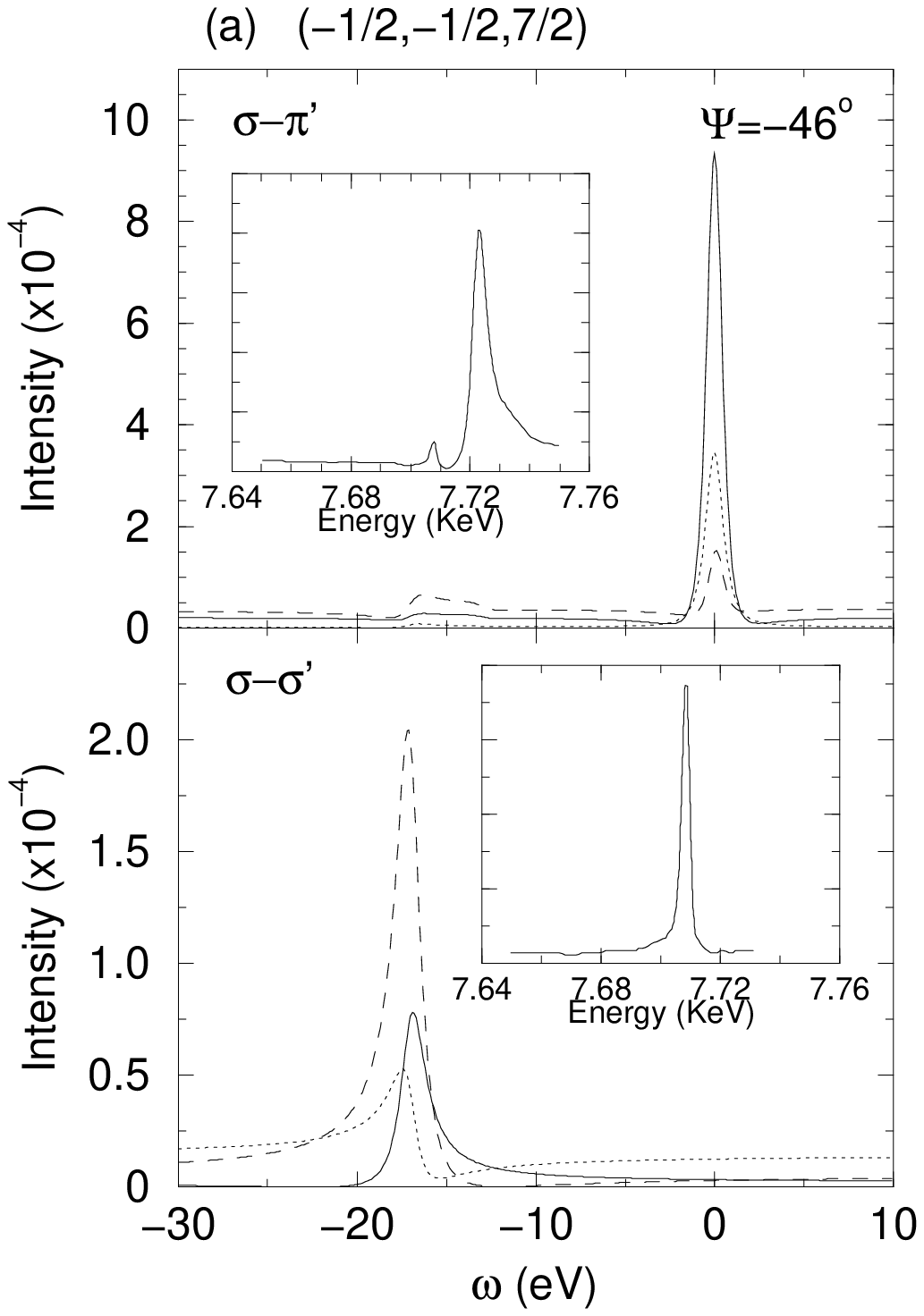,width=7cm}
\epsfile{file=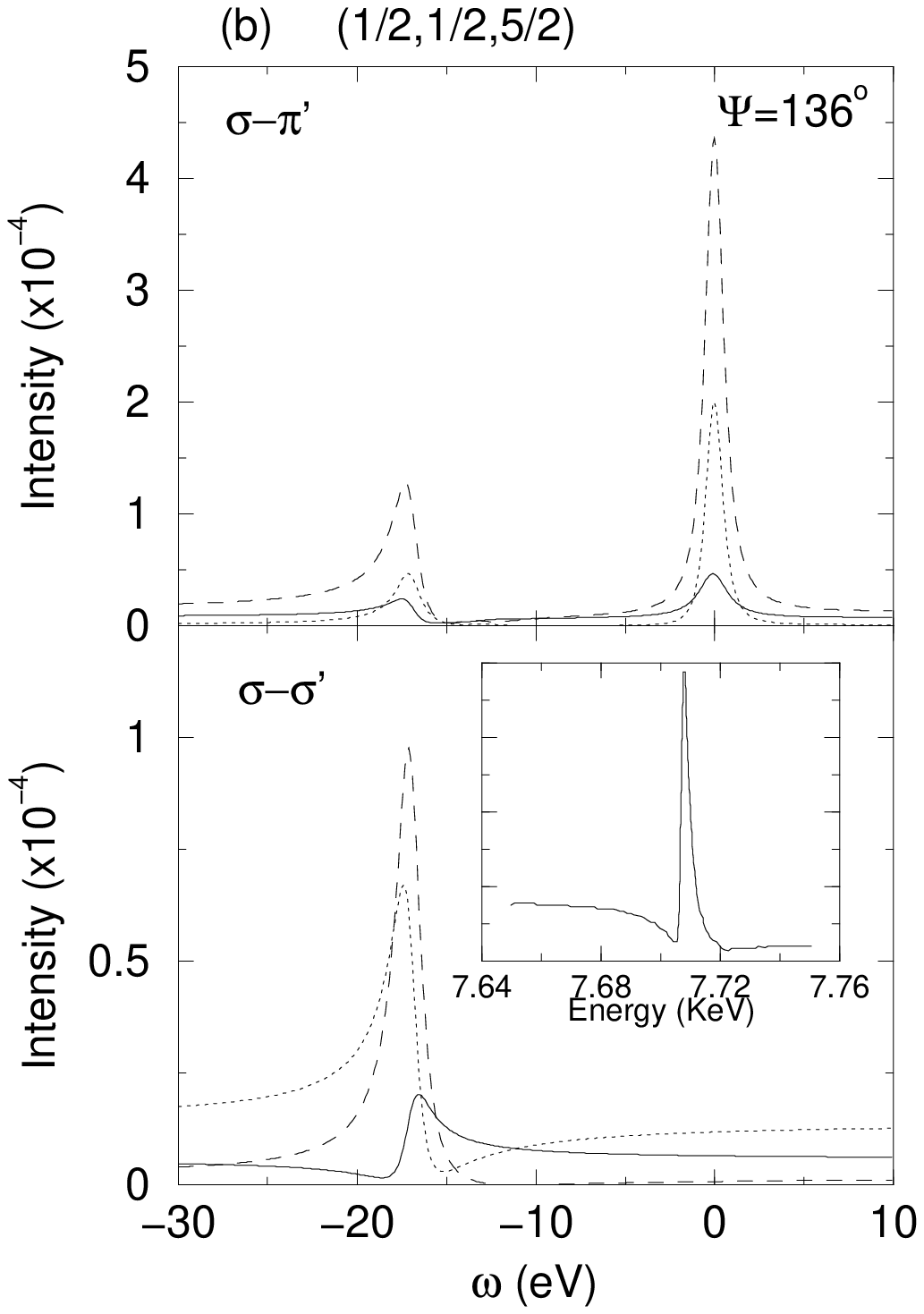,width=7cm}
\caption{
Magnetic scattering intensity as a function of photon energy
(a) for $\psi=-46^\circ$, ${\bf G}=(-\frac{1}{2},-\frac{1}{2},\frac{7}{2})$, 
and (b) for $\psi=136^\circ$, ${\bf G}=(\frac{1}{2},\frac{1}{2},\frac{5}{2})$.
The solid, dotted and broken lines represent the intensities from
domains $S_1$, $S_2$, and $S_3$
in the same $K$ domain with ${\bf Q}=(\frac{1}{2},\frac{1}{2},\frac{1}{2})$.
The zero of energy is at the main peak.
The upper and lower panels show the intensity for the $\sigma\to\pi'$
channel, and that for the $\sigma\to\sigma'$ channel, respectively.
The insets show the experimental curves of ref.~3.
The curve in the inset on the top panel in (a) is after absorption
correction.
}\label{fig.spec1}
\end{center}
\end{figure}

Figure \ref{fig.spec1} (b) shows the calculated intensity 
as a function of photon energy 
for ${\bf G}=(\frac{1}{2},\frac{1}{2},\frac{5}{2})$.
The azimuthal angle is set to be $\psi=136^\circ$, such that
the $(1,0,0)$ crystal axis lies on the scattering plane.
(This correspond to the experimental situation that the $(0,1,0)$ crystal
axis lies on the scattering vector.)
For the $\sigma\to\sigma'$ channel (lower panel), only the pre-edge peak
is found, in agreement with the experiment.
The anti-resonant dip is found below the pre-edge peak for domain $S_1$. 
Such a dip is actually observed in the experiment (see the inset).
For the $\sigma\to\pi'$ channel (upper panel), the main peak is found 
in addition to the pre-edge peak, but its intensity is much smaller
than that for ${\bf G}=(-\frac{1}{2},-\frac{1}{2},\frac{7}{2})$.
As already pointed out,\cite{Neubeck} the intensity of the main peak is
the smallest for domain $S_1$.
Experimentally, the main peak is not clearly seen.
Combining this fact to the spectral shape of the pre-edge peak,
we speculate that the experimental spectrum comes from domain $S_1$.

\begin{figure}
\begin{center}
\leavevmode
\epsfile{file=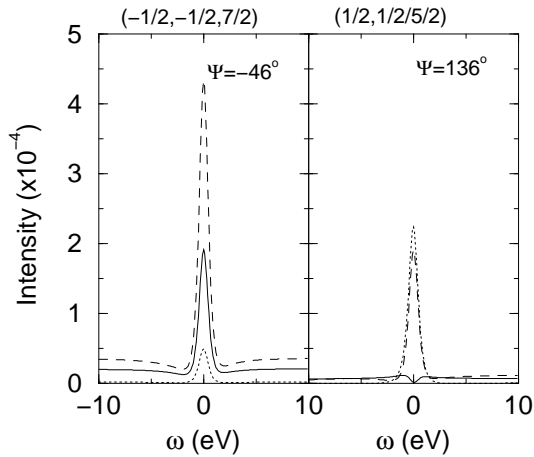,width=7cm}
\caption{
Magnetic scattering intensity for the $\sigma\to\pi'$ channel
in the energy region of main peak, calculated with neglecting
the intraatomic exchange interaction between the $4p$ and $3d$ states.
The solid, dotted and broken lines represent the intensities from
domains $S_1$, $S_2$ and $S_3$.
}\label{fig.spec2}
\end{center}
\end{figure}
Figure \ref{fig.spec2} shows the RXMS intensity for the $\sigma\to\pi'$
channel in the energy region of main peak, which is calculated with neglecting 
the intraatomic exchange interaction between the $4p$ and $3d$ states.
The intensities are nearly half of those shown in Fig.~\ref{fig.spec1},
indicating that the contribution of the intraatomic exchange interaction
is comparable to that of the $p$-$d$ mixing to the $3d$ states of
neighboring Co atoms.
Note that the latter contribution increases according to roughly
the square of the strength of the $p$-$d$ mixing.
It is highly probable that the effect of the intraatomic exchange
interaction is suppressed by a band formation of the $4p$ states,
since the energy splittings caused by the exchange interaction are
averaged out at the $A$ and $B$ sublattices in the states with small
momenta. For making clear this point, however, it is necessary to
perform a band calculation, which has not benn completed.

\begin{figure}
\begin{center}
\leavevmode
\epsfile{file=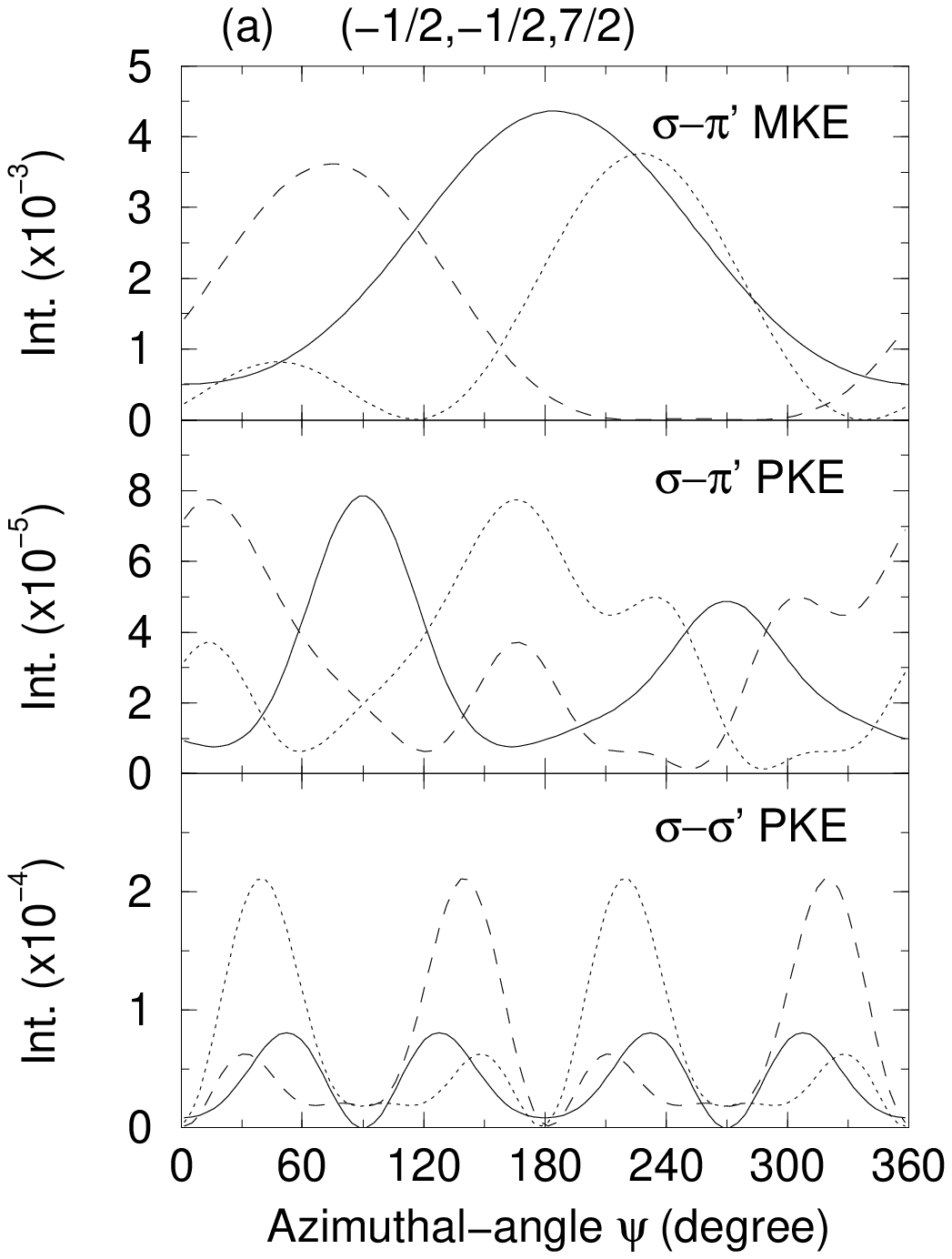,width=7cm}
\epsfile{file=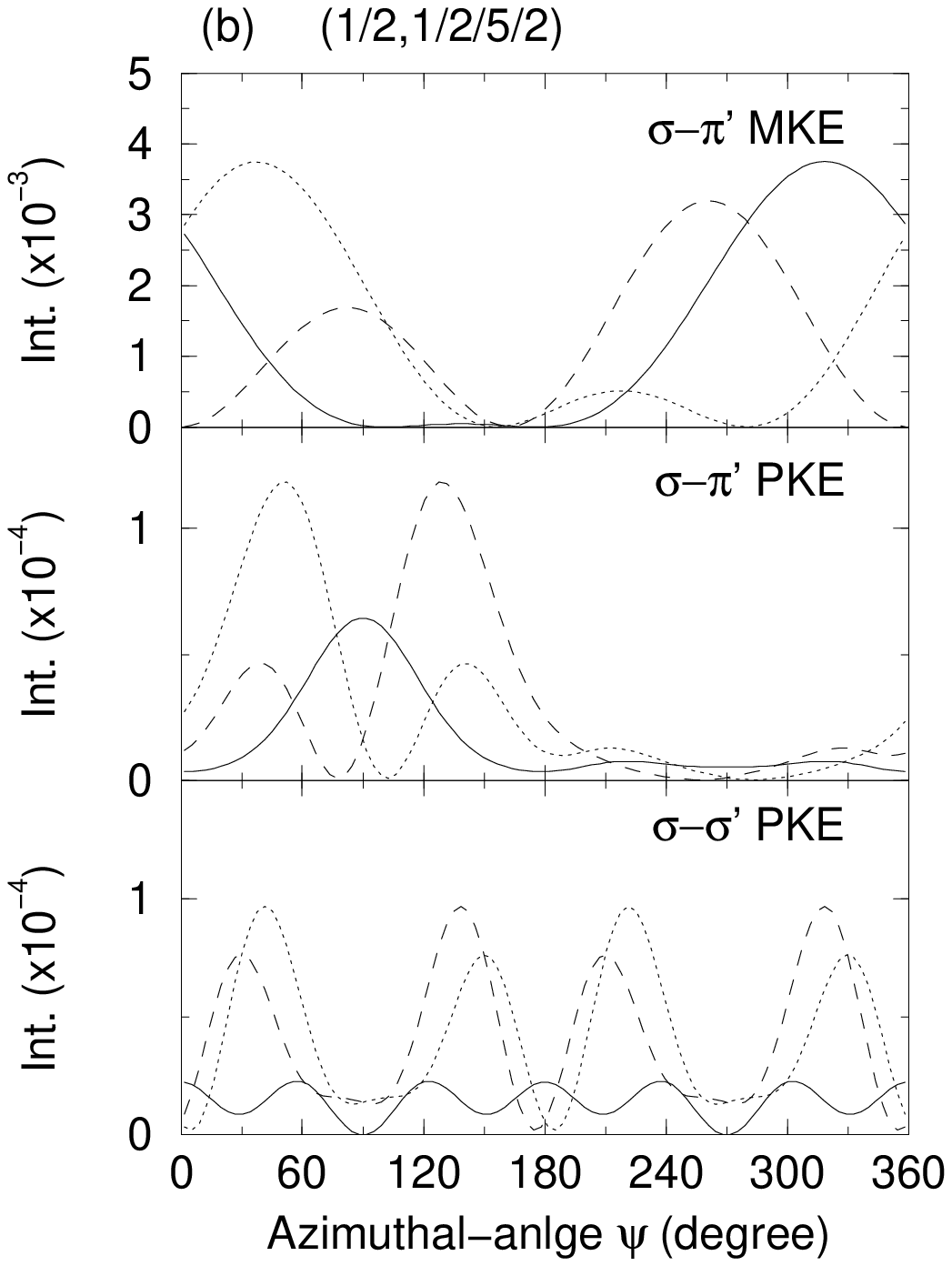,width=7cm}
\caption{
Magnetic scattering intensity as a function of azimuthal angle
$\psi$; (a) for ${\bf G}=(-\frac{1}{2},-\frac{1}{2},\frac{7}{2})$, 
and (b) for ${\bf G}=(\frac{1}{2},\frac{1}{2},\frac{5}{2})$.
The upper panel is for the main $K$-edge (MKE) peak, and
the middle and lower panels are for the pre $K$-edge (PKE) peak.
The solid, dotted and broken lines represent the intensities from
domains $S_1$, $S_2$ and $S_3$.
}\label{fig.azim}
\end{center}
\end{figure}

Figure \ref{fig.azim} shows the calculated intensity 
as a function of the azimuthal angle for
${\bf G}=(-\frac{1}{2},-\frac{1}{2},\frac{7}{2})$ and
$(\frac{1}{2},\frac{1}{2},\frac{5}{2})$. 
We find rather complicated behavior; for example,
for the $\sigma\to\pi'$ channel 
at ${\bf G}=(\frac{1}{2},\frac{1}{2},\frac{5}{2})$, 
the main peak grows considerably while the pre-edge peak shrinks, 
with increasing values of $\psi$.
One reason for such complicated behavior is that the scattering vector
is not along the crystal axes.
The experimental data are not available to make comparison.

\section{Concluding Remarks}

We have analyzed the recent experiment for the RXMS intensity for CoO 
using a cluster model.
The spins and orbitals are ordered with the same wavevector
under strong spin-orbit coupling.
The point different from the case of ordinary orbital order is that
the scattering tensor becomes antisymmetric. 
The cluster calculation gives a reasonable description for the RXMS intensity.
We have found that the $4p$ states of Co are modified not only by
the intraatomic exchange coupling to the $3d$ states  
but also by the orbital moment of the $3d$ states of neighboring Co atoms 
through the $p$-$d$ mixing.
Thereby the orbital moment is induced in the $4p$ states, and
a considerable intensity is brought about through the dipole process.
The $p$-$d$ mixing mechanism reminds us of the magnetic circular dichroism (MCD)
for the $K$-edge absorption in the ferromagnetic metals Fe, Co, and Ni,
\cite{Schutz} where the MCD intensity comes mainly from 
the orbital moment induced in the $4p$ states through the mixing to
the $3d$ states of neighboring Co atoms.\cite{Igarashi1,Igarashi2}

We have also found the pre-edge peak in the RXMS intensity.
It originates from the quadrupole process.
This contrasts with the pre-edge peak in the resonant scattering 
for the orbital order in LaMnO$_3$, which is predicted to be generated 
by the dipole process.\cite{Takahashi2}
We have also calculated the azimuthal angle dependence
of the RXMS intensity, which shows rather complicated behavior.
We are waiting for future experiments to make comparison.

In this paper, we have used a simple cluster model for analyzing 
qualitatively the experimental data. 
Since the $4p$ states are considerably extended in space,
it is much better to treat them as a band.
We speculate that the effect of the intraatomic exchange interaction
is suppressed by a band formation of the $4p$ states.
A study based on the band calculation is now under progress.

\section*{Acknowledgment}
We would like to thank W. Neubeck for kindly informing
the experimental values for the $\omega$, $\chi$, and $\phi$ angles 
in a 4-circle diffractometer.
We also acknowledge T. Jo and J. Mizuki for valuable discussions.
This work was partially supported by 
a Grant-in-Aid for Scientific Research in Priority Area,
``Novel Quantum Phenomena in Transition Metal Oxides
-Spin$\cdot$Charge$\cdot$Orbital Coupled Systems-"
No. 407, from the Ministry of Education, Science, Sports and Culture.

\appendix
\section{Geometrical Factors}

We outline the derivation of the geometrical factors in the coordinate frame 
fixed to the crystal axes and summarize the result here.

\subsection{Dipole Transition}

In the coordinate frame $(x',y',z')$ that the $z'$ axis is the propagation
direction of photon, the dipole operators for {\em circular} polarization
are given by
\begin{equation}
 \frac{1}{\sqrt{2}}(x'\pm iy') = \mp\left(\frac{4\pi}{3}\right)^{1/2}
                                  Y_{1\pm 1}(\theta',\varphi')r',
\end{equation}
where $r'$, $\theta'$, $\varphi'$ are corresponding polar coordinates.
We represent these operators in the coordinate frame $(x'',y'',z'')$
that the $z''$ axis is along the scattering vector ${\bf G}$ and
$y''$ axis is contained in the scattering plane (Fig.~\ref{fig.geom}).
This is done by the relation,\cite{Rose}
\begin{equation}
 Y_{\ell m'}(\theta',\varphi') = \sum_{m} D^{(\ell)}_{mm'}(\alpha,\beta,\gamma)
                                         Y_{\ell m}(\theta'',\varphi''),
\label{eq.rotation}
\end{equation}
where $\ell=1$, $m'=\pm 1$ with the Euler angles 
$\alpha$, $\beta$, $\gamma$ being $\alpha=\pi/2$, $\beta=\pi/2+\theta$, 
and $\gamma=-\pi/2$ for incident photons, and being $\alpha=\pi/2$, 
$\beta=\pi/2-\theta$, and $\gamma=-\pi/2$ for scattered photons.
According to the rotation of the crystal by azimuthal angle $\psi$ around
the scattering vector ${\bf G}$, we replace
\begin{equation}
 Y_{1m}(\theta'',\varphi'') \longrightarrow 
 Y_{1m}(\theta'',\varphi''){\rm e}^{im\psi''}.
\end{equation}
Next we then express $Y_{1m}(\theta'',\varphi'')$'s in the coordinate
frame $(x''',y''',z''')$ fixed to the crystal axes with the help of
eq.~(\ref{eq.rotation}), with
$\alpha=-3\pi/4$, $\beta=\arcsin\sqrt{2/51}$, $\gamma=0$,
for ${\bf G}=(-\frac{1}{2},-\frac{1}{2},\frac{7}{2})$,
and with $\alpha=\pi/4$, $\beta=\arcsin\sqrt{2/27}$, $\gamma=0$,
for ${\bf G}=(\frac{1}{2},\frac{1}{2},\frac{5}{2})$.
Finally we change the expressions in the circularly polarized base into
those in the linearly polarized base.
Thus we obtain the geometrical factor for incident photon 
$P^\mu$ ($\mu=\sigma$ or $\pi$) as 
\begin{eqnarray}
 \left(P^\sigma\right)_1 &=& \pm(\cos\beta\cos\psi + \sin\psi)/\sqrt{2},
                              \nonumber\\
 \left(P^\sigma\right)_2 &=& \pm(\cos\beta\cos\psi - \sin\psi)/\sqrt{2},
                              \nonumber\\
 \left(P^\sigma\right)_3 &=& -\sin\beta\cos\psi,
\label{eq.dsigma} \\
 \left(P^\pi\right)_1 &=& \pm\left[\sin\theta(\cos\beta\sin\psi-\cos\psi)
                                +\cos\theta\sin\beta\right]/\sqrt{2},
                                \nonumber\\
 \left(P^\pi\right)_2 &=& \pm\left[\sin\theta(\cos\beta\sin\psi+\cos\psi)
                                +\cos\theta\sin\beta\right]/\sqrt{2},
                                \nonumber\\
 \left(P^\pi\right)_3 &=& (-\sin\theta\sin\beta\sin\psi
                                +\cos\theta\cos\beta).
\label{eq.dpi}
\end{eqnarray}
The upper and lower signs in eqs.~(\ref{eq.dsigma}) and (\ref{eq.dpi})
correspond to ${\bf G}=(-\frac{1}{2},-\frac{1}{2},\frac{7}{2})$ and
$(\frac{1}{2},\frac{1}{2},\frac{5}{2})$, respectively.
We also obtain the geometrical factor for scattered photon 
$P'^{\mu'}$ ($\mu'=\sigma'$ or $\pi'$) from eqs.~(\ref{eq.dsigma}) 
and (\ref{eq.dpi}) with replacing $\sin\theta$ by $-\sin\theta$.
Note that the scattering plane for $\psi=0$ contains
the $(1,-1,0)$ crystal axis in this definition of the coordinate frame.

\subsection{Quadrupole Transition}

In the coordinate frame $(x',y',z')$ that the $z'$ axis is the
propagation direction of photon, the quadrupole operators 
for {\em circular} polarization are given by
\begin{equation}
 \frac{1}{2}\frac{1}{\sqrt{2}}z'(x'\pm iy') =
    \mp\left(\frac{\pi}{15}\right)^{1/2}Y_{2\pm 1}(\theta',\varphi').
\end{equation}
We represent these operators in the coordinate frame fixed to the
crystal axes by following the same procedure as for the dipole transition.
We change the expressions in the circularly polarized base into those 
in the linearly polarized base. Thus we obtain
the geometrical factor for incident photon $Q^\mu$ ($\mu=\sigma$ or $\pi$)
as
\begin{eqnarray}
 \left(Q^\sigma\right)_1 &=& 
    -\cos\theta\cos\beta\cos 2\psi
    -\sin\theta\sin\beta\sin\psi,
    \nonumber\\
 \left(Q^\sigma\right)_2 &=& 
    (\sqrt{3}/2)\cos\theta\sin^2\beta\sin 2\psi
    +(\sqrt{3}/2)\sin\theta\sin 2\beta\cos\psi,
    \nonumber\\
 \left(Q^\sigma\right)_3 &=& 
    \mp\cos\theta\sin\beta(\cos\beta\sin 2\psi+\cos 2\psi)/\sqrt{2}\nonumber\\
   &\mp&\sin\theta(\cos 2\beta\cos\psi-\cos\beta\sin\psi)/\sqrt{2},
    \nonumber\\
 \left(Q^\sigma\right)_4 &=& 
    \mp\cos\theta\sin\beta(\cos\beta\sin 2\psi-\cos 2\psi)/\sqrt{2}\nonumber\\
   &\mp&\sin\theta(\cos 2\beta\cos\psi+\cos\beta\sin\psi)/\sqrt{2},
    \nonumber\\
 \left(Q^\sigma\right)_5 &=& 
    \cos\theta[1-(1/2)\sin^2\beta]\sin 2\psi
    -(1/2)\sin\theta\sin 2\beta\cos\psi,
\label{eq.qsigma} \\
 \left(Q^\pi\right)_1 &=& 
    -(1/2)\sin 2\theta\cos\beta\sin 2\psi
    -\cos 2\theta\sin\beta\cos\psi,
    \nonumber\\
 \left(Q^\pi\right)_2 &=& 
    -(\sqrt{3}/4)\sin 2\theta\sin^2\beta\cos 2\psi
    -(\sqrt{3}/2)\cos 2\theta\sin 2\beta\sin\psi\nonumber\\
   &-&(\sqrt{3}/2)\sin 2\theta[1-(3/2)\sin^2\beta],
    \nonumber\\
 \left(Q^\pi\right)_3 &=& 
    \pm(1/2\sqrt{2})\sin 2\theta\sin\beta(\cos\beta\cos 2\psi-\sin 2\psi)
    \nonumber\\
   &\pm&\cos 2\theta(\cos 2\beta\sin\psi+\cos\beta\cos\psi)/\sqrt{2}\nonumber\\
   &\mp&(3/4\sqrt{2})\sin 2\theta\sin 2\beta,
    \nonumber\\
 \left(Q^\pi\right)_4 &=& 
     \pm(1/2\sqrt{2})\sin 2\theta\sin\beta(\cos\beta\cos 2\psi+\sin 2\psi)
    \nonumber\\
   &\pm&\cos 2\theta(\cos 2\beta\sin\psi-\cos\beta\cos\psi)/\sqrt{2}\nonumber\\
   &\mp&(3/4\sqrt{2})\sin 2\theta\sin 2\beta,
    \nonumber\\
 \left(Q^\pi\right)_5 &=& 
    -(1/2)\sin 2\theta[1-(1/2)\sin^2\beta]\cos 2\psi
    \nonumber\\
   &+&(1/2)\cos 2\theta\sin 2\beta\sin\psi
    -(3/4)\sin 2\theta\sin^2\beta,
\label{eq.qpi}
\end{eqnarray}
where $\beta=\arcsin\sqrt{2/51}$ for 
${\bf G}=(-\frac{1}{2},-\frac{1}{2},\frac{7}{2})$,
and $\beta=\arcsin\sqrt{2/27}$ for 
${\bf G}=(\frac{1}{2},\frac{1}{2},\frac{5}{2})$.
The upper and lower signs in eqs.~(\ref{eq.qsigma}) and (\ref{eq.qpi})
correspond to ${\bf G}=(-\frac{1}{2},-\frac{1}{2},\frac{7}{2})$ and
$(\frac{1}{2},\frac{1}{2},\frac{5}{2})$.
We also obtain the geometrical factor for scattered photon 
$Q'^{\mu'}$ ($\mu'=\sigma'$ or $\pi'$) 
from eqs.~(\ref{eq.qsigma}) and (\ref{eq.qpi})
with replacing $\sin\theta$ and $\sin 2\theta$ by $-\sin\theta$ 
and $-\sin 2\theta$, respectively.


\begin{thebibliography}{99}
\bibitem{com1} In general, ${\bf L}({\bf G})$ is similar
               to the Fourier transform of the orbital angular momentum 
               density, but is not exactly reduced to that,
               except for ${\bf G}=0$. See ref.~19 for the explicit form.
\bibitem{com2} Setting the energy of the state $p'_0$ as the origin,
               we estimate
               $\epsilon_1=-\frac{3}{15}G^1+\frac{6}{245}G^3$,
               $\epsilon_2=-\frac{15}{245}G^3$, with $G^1$ and $G^3$ being 
               the Slater integrals between the $4p$ and $3d$ states.
               We evaluate $G^1$ and $G^2$ within the ion radius as
               $G^1\sim G^3=0.4$ eV, since the $4p$ states lose their 
               meaning outside the ion radius, and further multiply 
               a factor $0.8$ to those values to take account of 
               solid-state effect.
\end{thebibliography}
\end{document}